# Analytical Solution for the Anisotropic Rabi Model: Effects of Counter-Rotating Terms

Guofeng Zhang(张国锋)[1,2,3,*], Hanjie Zhu(朱汉杰)[1]

[1]*Department of Physics, School of Physics and Nuclear Energy Engineering,*
*Beihang University, Xueyuan Road No. 37, Beijing 100191, China*

[2]*State Key Laboratory of Theoretical Physics(SKLTP), Institute of Theoretical Physics,*
*Chinese Academy of Sciences, Beijing, 100190, China*

[3]*State Key Laboratory of Low-Dimensional Quantum Physics, Tsinghua University, Beijing 100084, China*

**The anisotropic Rabi model, which was proposed recently, differs from the original Rabi model: the rotating and counter-rotating terms are governed by two different coupling constants. This feature allows us to vary the counter-rotating interaction independently and explore the effects of it on some quantum properties. In this paper, we eliminate the counter-rotating terms approximately and obtain the analytical energy spectrums and wavefunctions. These analytical results agree well with the numerical calculations in a wide range of the parameters including the ultrastrong coupling regime. In the weak counter-rotating coupling limit we find out that the counter-rotating terms can be considered as the shifts to the parameters of the Jaynes-Cummings model. This modification shows the validness of the rotating-wave approximation on the assumption of near-resonance and relatively weak coupling. Moreover, the analytical expressions of several physics quantities are also derived, and the results show the break-down of the U(1)-symmetry and the deviation from the Jaynes-Cummings model.**

The Rabi model[1], one of the simplest model which deals with the matter-light interaction, describes the interaction of a two-level atom with a single bosonic mode. This ubiquitous model is applied to a great variety of physical systems, such as microwave and optical cavity quantum electrodynamics (QED)[2], ion traps[3], quantum dots, and circuit QED[4-7]. The Rabi model Hamiltonian reads

$$H_R = \omega a^+ a + \Omega \sigma_z + g\sigma_x(a^+ + a), \qquad (1)$$

where $a^+$ and $a$ are creation and annihilation operators of the mode with frequency $\omega$, $\sigma_{x,y,z}$ are the Pauli spin operators associated to a two-level system with transition frequency $2\Omega$, and $g$ is the interaction strength. For simplicity, we already take the unit of $\hbar = 1$.

[*] Correspondence and requests for materials should be addressed to G.Z.(gf1978zhang@buaa.edu.cn)

Although much attention has been paid over the last decades, until now the exact analytical solution of Rabi model is still lacking. To overcome this problem, the most used analytical method is the rotating-wave approximation (RWA), where the counter-rotating terms $a^+\sigma_+ + a\sigma_-$ is neglected. This approximation leads to the Jaynes-Cummings (JC) model[8], which can be solved exactly. In the regime of near resonance and relatively weak coupling, the JC model captures the quantum dynamics of Rabi model successfully. However, solid-state systems such as superconductor systems have allowed the coupling strength to reach the ultrastrong coupling regime $(g/\omega \sim 0.1)^6$. In this regime, the counter-rotating terms are no longer negligible. This leads to the break-down of RWA and the system can only be described by the Rabi model. Consequently, a variety of method has been developed to go beyond the RWA[9-18].

Recently, a generalization of the Rabi model called the anisotropic Rabi model was proposed[19,20]. The Hamiltonian of anisotropic Rabi model can be written as:

$$H = \omega a^+ a + \Omega \sigma_z + g(\sigma_+ a + \sigma_- a^+) + g'(\sigma_- a + \sigma_+ a^+). \qquad (2)$$

Here $\sigma_\pm = (\sigma_x \pm i\sigma_y)/2$, $g$ and $g'$ denote the coupling strength of the rotating terms $\sigma_+ a + \sigma_- a^+$ and counter-rotating terms $\sigma_- a + \sigma_+ a^+$ respectively. In this model, the coupling strength of the rotating wave interaction is different from that of the counter-rotating wave interaction. The anisotropic Rabi model includes the JC model ($g' = 0$) and the original Rabi model ($g' = g$), and has the applications to different physical fields, such as quantum optics, solid state physics, and mesoscopic physics [19]. Besides these practical applications, this model is a significant key for us to understand the characters of the counter-rotating terms. The independence of two coupling constants allows us to explore the effects of the counter-rotating terms, while in the original Rabi model it is hard to separate the influences of two types of interaction terms. Although this model can be solved exactly[19] by the method originally developed by Braak[12], the results are strongly dependent of the composite transcendental function defining through its power series in the interacting strength and the frequency, and are difficult to extract the fundamental physics of the model.

**Elimination of the counter-rotating terms**

We now begin to eliminate the counter-rotating terms. The Hamiltonian of the anisotropic Rabi model has the equivalent form

$$H = \omega a^+ a + \Omega \sigma_z + g_1 \sigma_x (a + a^+) + i g_2 \sigma_y (a^+ - a), \qquad (3)$$

where $g_1 = (g + g')/2$ and $g_2 = (g' - g)/2$. We note that the anisotropic Rabi model also possesses $\mathbb{Z}_2$-symmetry. If we define a parity operator $P = \sigma_z e^{i\pi a^+ a}$, it is obvious that $[H, P] = 0$. Thus the state space can be decomposed into two subspaces $\mathcal{H}_\pm$, where the parity operator $P$ has eigenvalues $\pm 1$ respectively.

Now we apply the unitary transformation $U = exp[\lambda \sigma_x (a^+ - a)]$. Here $\lambda$ is the dimensionless parameter determined by the later calculations. By performing this transformation, the Hamiltonian becomes $H' = UHU^+ = H_1 + H_2 + H_3 + H_4$, where

$$H_1 = \omega a^+ a - \lambda \omega \sigma_x (a^+ + a) + \omega \lambda^2, \qquad (4)$$
$$H_2 = g_1 [\sigma_x (a^+ + a) - 2\lambda], \qquad (5)$$
$$H_3 = \Omega \{\sigma_z cosh[2\lambda(a^+ - a)] - i\sigma_y sinh[2\lambda(a^+ - a)]\}, \qquad (6)$$

$$H_4 = i g_2 (a^+ - a) \{\sigma_y cosh[2\lambda(a^+ - a)] + i\sigma_z sinh[2\lambda(a^+ - a)]\}. \qquad (7)$$

The operator $cosh[2\lambda(a^+ - a)]$ can be expanded as:

$$cosh[2\lambda(a^+ - a)] = \sum_{m,n} \langle m|cosh[2\lambda(a^+ - a)]|n\rangle |m\rangle\langle n|. \qquad (8)$$

The matrix elements in Eq. (8) can be calculated directly by using the formula

$$\langle N|e^{\lambda(a^+ - a)}|M\rangle = e^{-\frac{\lambda^2}{2}} \lambda^{N-M} \sqrt{\frac{M!}{N!}} L_M^{N-M}(\lambda^2), \qquad (M \leq N). \qquad (9)$$

From Eq. (9) we know that the remote matrix elements in Eq. (8) are the high-order terms of $\lambda$. When $\lambda$ is much smaller than 1, these terms can be discarded and only the diagonal elements are retained. Then we have:

$$\cosh[2\lambda(a^+ - a)] = \sum_n \langle n|\cosh[2\lambda(a^+ - a)]|n\rangle\, |n\rangle\langle n|. \quad (10)$$

This approximation can be interpreted as neglecting the multi-photon process in the effective Hamiltonian $H'$. When $\lambda \ll 1$, the multi-photon process is relatively weak and can be ignored. Following the same approximation procedure:

$$\sinh[2\lambda(a^+ - a)] = \sum_n \langle n|\sinh[2\lambda(a^+ - a)]|n-1\rangle(|n\rangle\langle n-1| - |n-1\rangle\langle n|), \quad (11)$$

$$(a^+ - a)\cosh[2\lambda(a^+ - a)] = \sum_n \langle n|(a^+ - a)\cosh[2\lambda(a^+ - a)]|n-1\rangle(|n\rangle\langle n-1| - |n-1\rangle\langle n|), \quad (12)$$

$$(a^+ - a)\sinh[2\lambda(a^+ - a)] = \sum_n \langle n|(a^+ - a)\sinh[2\lambda(a^+ - a)]|n\rangle |n\rangle\langle n|. \quad (13)$$

As a result, the effective Hamiltonian reduces to the form

$$H' = \omega a^+ a + \omega\lambda^2 - 2g_1\lambda + (g_1 - \lambda\omega)\sigma_x(a^+ + a)$$
$$+ \sigma_z \sum_n G_n |n\rangle\langle n| - i\sigma_y \sum_n R_n (|n\rangle\langle n-1| - |n-1\rangle\langle n|), \quad (14)$$

where

$$G_n = \Omega\langle n|\cosh[2\lambda(a^+ - a)]|n\rangle - g_2\langle n|(a^+ - a)\sinh[2\lambda(a^+ - a)]|n\rangle, \quad (15)$$

$$R_n = \Omega\langle n|\sinh[2\lambda(a^+ - a)]|n-1\rangle - g_2\langle n|(a^+ - a)\cosh[2\lambda(a^+ - a)]|n-1\rangle. \quad (16)$$

The expressions of $G_n$ and $R_n$ can be derived straightforwardly by using the Eq. (9)

$$R_n = \frac{2\Omega}{\sqrt{n}}\lambda e^{-2\lambda^2}L^1_{n-1}(4\lambda^2) - g_2\sqrt{n}e^{-2\lambda^2}L_{n-1}(4\lambda^2) + \frac{4g_2}{\sqrt{n}}e^{-2\lambda^2}\lambda^2 L^2_{n-1}(4\lambda^2), \quad (17)$$

$$G_n = \begin{cases} \Omega e^{-2\lambda^2} + 2g_2\lambda e^{-2\lambda^2} & (n=0) \\ \Omega e^{-2\lambda^2}L_n(4\lambda^2) + 2g_2\lambda e^{-2\lambda^2}[L^1_{n-1}(4\lambda^2) + L^1_n(4\lambda^2)] & (n \geq 1) \end{cases}. \quad (18)$$

Here $L_n(y)$ is the Laguerre polynomial and $L^i_n(y)$ is the associated Laguerre polynomial.

Since $\sigma_x = \sigma_+ + \sigma_-$, $-i\sigma_y = -\sigma_+ + \sigma_-$, the effective Hamiltonian becomes

$$H' = \omega a^+ a + \sigma_z \sum_n G_n |n\rangle\langle n| + \omega\lambda^2 - 2g_1\lambda + \sum_n (\sigma_+|n-1\rangle\langle n| + \sigma_-|n\rangle\langle n-1|)\left((g_1 - \lambda\omega)\sqrt{n} + R_n\right)$$
$$+ \sum_n (\sigma_-|n-1\rangle\langle n| + \sigma_+|n\rangle\langle n-1|)\left((g_1 - \lambda\omega)\sqrt{n} - R_n\right). \quad (19)$$

Now we have transformed the original Hamiltonian into a form in which the counter-rotating terms can be adjusted by tuning the dimensionless parameter $\lambda$.

**Energy Spectrums and Wavefunctions**

We now begin to derive the energy spectrums of the anisotropic Rabi model. In Eq. (19), the counter-rotating terms $\sigma_-|n-1\rangle\langle n| + \sigma_+|n\rangle\langle n-1|$ can be eliminated if the dimensionless parameter $\lambda$ is chosen as

$$\lambda = \lambda_n, \qquad (g_1 - \lambda_n\omega)\sqrt{n} - R_n = 0. \tag{20}$$

When $n=1$, the elimination of the term $\sigma_-|0\rangle\langle 1| + \sigma_+|1\rangle\langle 0|$ makes $\{|-z,0\rangle\}$ becomes an invariant subspace of $H'$. Then the ground-state energy can be written as

$$E_G = \langle -z,0|H'|-z,0\rangle = \omega\lambda_1^2 - 2g_1\lambda_1 - \Omega e^{-2\lambda_1^2} - 2g_2\lambda e^{-2\lambda_1^2}. \tag{21}$$

The ground-state wavefunction can be carry out immediately as

$$|\phi_G\rangle = e^{-\lambda_1\sigma_x(a^+-a)}|-z,0\rangle. \tag{22}$$

Since $\lambda_{n+1} \neq \lambda_{n-1}$, it seems unable to eliminate the terms $\sigma_-|n-2\rangle\langle n-1| + \sigma_+|n-1\rangle\langle n-2|$ and $\sigma_-|n\rangle\langle n+1| + \sigma_+|n+1\rangle\langle n|$ simultaneously. But the numerical results show that $|\lambda_{n+1} - \lambda_n|$ and $|\lambda_{n-1} - \lambda_n|$ is much smaller than $\lambda_n$, so we can neglect the difference between $\lambda_{n\pm1}$ and $\lambda_n$, and choose $\lambda = \lambda_n$, then $\{|+z,n-1\rangle, |-z,n\rangle\}$ becomes an invariant subspace of $H'$. When written in the basis of the states $|+z,n-1\rangle, |-z,n\rangle$, the Hamiltonian becomes

$$H_n = \begin{pmatrix} (n-1)\omega + \omega\lambda_n^2 - 2g_1\lambda_n + G_{n-1} & 2R_n \\ 2R_n & n\omega + \omega\lambda_n^2 - 2g_1\lambda_n - G_n \end{pmatrix}. \tag{23}$$

And the excited-state energy can be given by

$$E_{n,\pm} = \left(n - \frac{1}{2}\right)\omega + \omega\lambda_n^2 - 2g_1\lambda_n + \frac{G_{n-1} - G_n}{2} \pm \sqrt{\left[\frac{-\omega + G_{n-1} + G_n}{2}\right]^2 + 4R_n^2}. \tag{24}$$

The eigenvectors for $H_n$ are given by

$$|E_{n-}\rangle = cos\theta_n|+z,n-1\rangle + sin\theta_n|-z,n\rangle, \tag{25}$$
$$|E_{n+}\rangle = -sin\theta_n|+z,n-1\rangle + cos\theta_n|-z,n\rangle, \tag{26}$$

where

$$tan2\theta_n = \frac{2R_n}{E_{n-} - E_{n+}}. \tag{27}$$

Then we obtain the excited-state wavefunctions:

$$|\phi_{n-}\rangle = e^{-\lambda_n\sigma_x(a^+-a)}|E_{n-}\rangle = e^{-\lambda_n\sigma_x(a^+-a)}(cos\theta_n|+z,n-1\rangle + sin\theta_n|-z,n\rangle), \tag{28}$$
$$|\phi_{n+}\rangle = e^{-\lambda_n\sigma_x(a^+-a)}|E_{n+}\rangle = e^{-\lambda_n\sigma_x(a^+-a)}(-sin\theta_n|+z,n-1\rangle + cos\theta_n|-z,n\rangle). \tag{29}$$

Since the anisotropic Rabi model also possesses $\mathbb{Z}_2$-symmetry, the inexistence of level crossings within the subspaces $\mathcal{H}_\pm$ allows us to label each eigenstate with two quantum numbers $|n_0, n_1\rangle$, the same as the case of the Rabi model in the article [12]. The parity quantum number $n_0$ takes the values $\pm 1$ which corresponds to the subspaces $\mathcal{H}_\pm$. Within each subspace the states are labeled with the quantum number $n_1 = 0,1,2,\ldots$. Using this notation, our analytical state can be labelled as

$$|\phi_G\rangle \to |-1,0\rangle,$$

$$|\phi_{2m,-}\rangle \to |-1,2m-1\rangle, \quad |\phi_{2m,+}\rangle \to |-1,2m\rangle,$$

$$|\phi_{2m-1,-}\rangle \to |+1,2m-2\rangle, \quad |\phi_{2m-1,+}\rangle \to |+1,2m-1\rangle. \quad (30)$$

Fig.1 shows the lowest part of the energy spectrum from our analytical results for $g' = 2g$ and $g' = g/2$ respectively. For comparison purposes, the energy spectrum obtained from the numerical calculations is also shown. In this figure we find that the analytical energy spectrum agrees well with the numerical results both for $g' > g$ and $g' < g$ in the regime $g \leq 0.5$. In the case of $g = g'$, the model returns to the original Rabi model and has been discuss by the article [11].

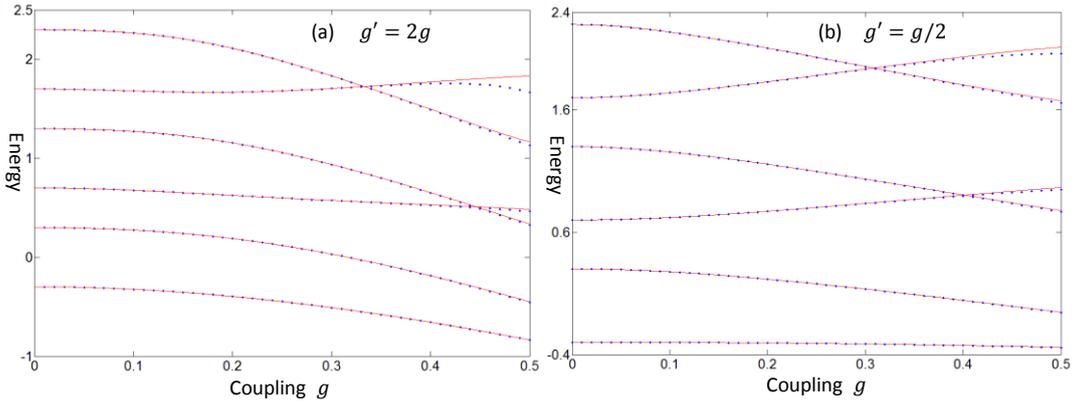

Figure 1: The lowest part of the energy spectrum of the anisotropic Rabi model as a function of the interaction strength $g$ for $\omega = 1$, $\Omega = 0.3$ and (a) $g' = 2g$, (b) $g' = g/2$. In all figures, the solid lines and dotted lines correspond to the analytical results and numerical results respectively.

It is necessary to discuss the valid parameter regime of our approximation. Our approximation procedures in Eqs. (10-13) require the dimensionless parameter $\lambda$ to be less than 1. When $\lambda$ approaches 1, this approximation is no longer valid and the analytical results start to fail considerably. In Fig.2, the dimensionless parameter $\lambda = \lambda_1$ as a function of the interaction strength $g$ and $g'$ are plotted. In the regime of $\lambda \leq 0.5$ our approximation results have a good agreement with the numerical results.

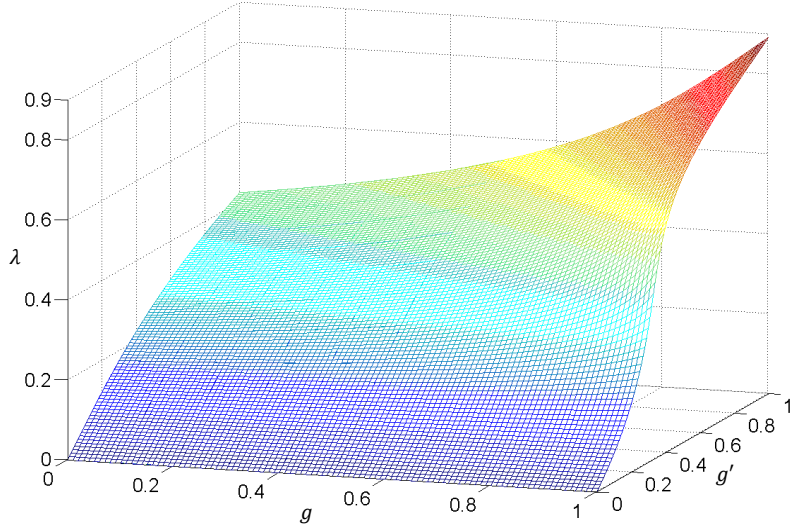

Figure 2: The dimensionless parameter $\lambda = \lambda_1$ as a function of the interaction strengths $g$ and $g'$ for $\omega = 1$, $\Omega = 0.3$.

**Limit of the weak counter-rotating coupling**

We now discuss a particular interesting and significant situation of $g' \ll \omega$, which corresponds to the weak counter-rotating coupling limit. When the anisotropic Rabi model return to the JC model ($g' = 0$), it is obvious that $\lambda = 0$. This indicates that we have $\lambda \ll 1$ when $g' \ll \omega$. This can also be confirmed in Fig.2.

In the previous discussion, we know that $\lambda$ satisfies the equation

$$0 = (g_1 - \lambda\omega)\sqrt{n} - R_n. \qquad (31)$$

Since $L_m^n(x) \approx \binom{m+n}{m}$ when $x \ll 1$, by neglecting the high-order terms of $\lambda$, the Eq. (31) reduces to the form $g_1 - \lambda\omega - 2\Omega\lambda + g_2 = 0$. It leads to the solution:

$$\lambda = \frac{g_1 + g_2}{\omega + 2\Omega} = \frac{g'}{\omega + 2\Omega}. \qquad (32)$$

Now we have obtained the analytical expression of $\lambda$ when $\lambda \ll 1$. Therefore, in the weak counter-rotating coupling limit, the dimensionless parameter $\lambda$ is proportional to $g'$, this reveals the physical meaning of $\lambda$. In the anisotropic Rabi model, $g'$ describes the '**absolute deviation**' from the JC model, which is the coupling strength of the counter-rotating wave interaction. Then we may regard $\lambda$ as the '**relative deviation**' from the JC model.

When $\lambda \ll 1$, we have

$$G_n = \Omega + 2g_2\lambda(2n + 1), \qquad R_n = (g_1 - \lambda\omega)\sqrt{n}. \qquad (33)$$

On substituting Eq. (33) into the effective Hamiltonian Eq. (19), we obtain

$$H' = \omega a^+ a + \omega \lambda^2 - 2g_1\lambda + \sigma_z \sum_n [\Omega + 2g_2\lambda(2n+1)]|n\rangle\langle n|$$

$$+ 2\sum_n (g_1 - \lambda\omega)\sqrt{n}(\sigma_+|n-1\rangle\langle n| + \sigma_-|n\rangle\langle n-1|), \quad (34)$$

which can be written as

$$H' = (\omega + \Delta\omega\sigma_z)a^+a + (\Omega + \Delta\Omega)\sigma_z + (g + \Delta g)(\sigma_+ a + \sigma_- a^+) + \Delta E, \quad (35)$$

where

$$\Delta\omega = 4g_2\lambda, \quad \Delta\Omega = 2g_2\lambda,$$
$$\Delta g = \frac{2\Omega - \omega}{\omega + 2\Omega}g', \quad \Delta E = -2g_1\lambda. \quad (36)$$

Here we have obtained a modified JC model with an additional term $\sigma_z a^+ a$. The counter-rotating terms have been considered as shifts to the parameters instead of being abandoned directly in the RWA. This is an improved approximation compares to the RWA.

This modification of JC model allows us to discuss the validness of RWA. In the regime of near resonance $2\Omega \approx \omega$, we have $\Delta g \approx 0$. Hence it is reasonable to ignore the shift to the coupling strength, so RWA captures the changes of atom-field interactions very good in this regime. When $g' \ll \omega$, such as the optical cavity with strong coupling, the shifts to the other parameters is relatively small, then the RWA becomes a successful approximation. When the system is under ultrastrong coupling $(g \sim 0.1\omega)$, the shifts to the other parameters become nonnegligible. Therefore, the effects of counter-rotating terms begin to appear and RWA fails to capture it, such as the Bloch-Siegert shift.

**The effects of counter-rotating terms on physics quantities**

Based on the energy spectrums and the wavefunctions, we are able to derive the corresponding physics quantities, and discuss the effects of counter-rotating terms on them. Firstly we calculate the Bloch-Siegert shift, which is the energy shift of the level transition due to the counter-rotating terms[21]. The Bloch-Siegert shift with the transition $E_{1-} \to E_G$ can be calculated immediately as

$$\delta = \Omega e^{-2\lambda^2}(1+2\lambda^2) - \Omega + 4g_2\lambda^3 e^{-2\lambda^2} + \sqrt{\left(\Omega - \frac{\omega}{2}\right)^2 + g^2}$$

$$-\sqrt{\left[-\frac{\omega}{2} + \Omega e^{-2\lambda^2}(1-2\lambda^2) + 4g_2\lambda e^{-2\lambda^2}(1-\lambda^2)\right]^2 + 4(g_1 - \omega\lambda)^2}. \quad (37)$$

It is obvious that $\delta = 0$ when the anisotropic Rabi model returns to the JC model ($\lambda = 0$). In Fig.3 we show the absolute value of the Bloch-Siegert shift with the transition $E_{1-} \to E_G$ as a function of $g$ and $g'$. In this figure, the analytical results agree perfectly with the numerical calculation even when $\lambda \sim 0.6$.

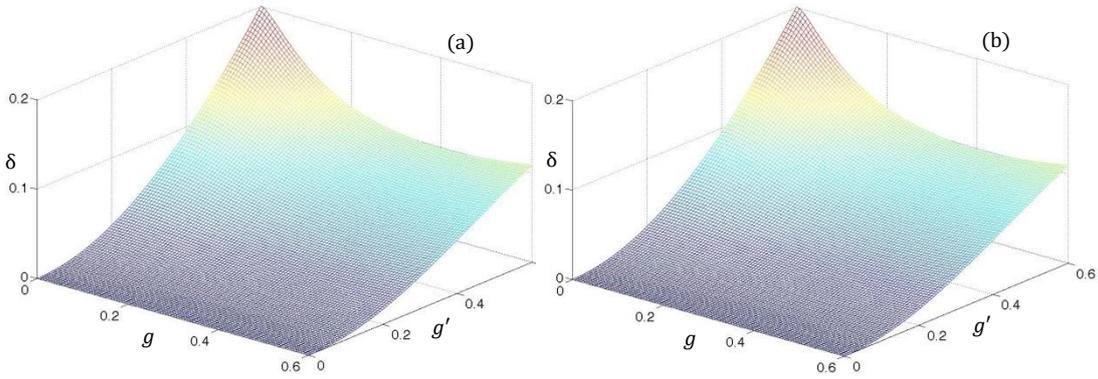

Figure 3: The absolute value of the Bloch-Siegert shift with the transition $E_{1-} \to E_G$ as a function of $g$ and $g'$ for $\omega = 1$, $\Omega = 0.3$. The part (a) represents the analytical results, while the part (b) corresponds to the numerical results.

The mean photon number $\langle a^+ a \rangle$ and $\langle \sigma_z \rangle$ can also be evaluated. For the ground state, they can be given by

$$\langle \phi_G | a^+ a | \phi_G \rangle = \lambda^2, \qquad \langle \phi_G | \sigma_z | \phi_G \rangle = -e^{-2\lambda^2}. \quad (38)$$

For the excited state, we have

$$\langle \phi_{n\pm} | a^+ a | \phi_{n\pm} \rangle = n - \frac{1}{2} + \lambda^2 \pm \left(\frac{1}{2}\cos 2\theta_n + 2\lambda\sqrt{n}\sin\theta_n \cos\theta_n\right), \quad (39)$$

$$\langle \phi_{n-} | \sigma_z | \phi_{n-} \rangle = \cos^2\theta_n e^{-2\lambda^2} L_{n-1}(4\lambda^2) - \sin^2\theta_n e^{-2\lambda^2} L_n(4\lambda^2) + \cos\theta_n \sin\theta_n \frac{4}{\sqrt{n}} \lambda e^{-2\lambda^2} L^1_{n-1}(4\lambda^2), \quad (40)$$

$$\langle \phi_{n+} | \sigma_z | \phi_{n+} \rangle = \sin^2\theta_n e^{-2\lambda^2} L_{n-1}(4\lambda^2) - \cos^2\theta_n e^{-2\lambda^2} L_n(4\lambda^2) - \cos\theta_n \sin\theta_n \frac{4}{\sqrt{n}} \lambda e^{-2\lambda^2} L^1_{n-1}(4\lambda^2). \quad (41)$$

These results enable us to evaluate the number of polaritons $N = a^+ a + \sigma_z/2 + \mathbb{I}/2$. In the JC model, $N$ is a conservation quantity and leads to the U(1)-symmetry. In the presence of the counter-rotating terms, the U(1)-symmetry is broken down to the $\mathbb{Z}_2$-symmetry and $N$ is no longer conserved. The mean number of polaritons of the ground state can be given by

$$\langle N \rangle_G = \lambda^2 - \frac{1}{2}e^{-2\lambda^2} + \frac{1}{2} = \lambda^4 + o(\lambda^5). \quad (42)$$

The Eq. (42) shows that $\langle N \rangle_G$ is increased with $\lambda$. This is due to the transition from $|-z, 0\rangle$ to the upper level caused by the counter-rotating terms. In the JC model, this kind of transition is forbidden since no counter-rotating terms. This transition makes the ground state becomes the superposition state of $|-z, 0\rangle$ and the other states. Therefore, the mean number of polaritons of the ground state is increased with the coupling strength of the counter-rotating terms.

The uncertainty $\Delta N$ of the ground state can also be given as

$$(\Delta N)^2 = \frac{3}{2}\lambda^2 e^{-2\lambda^2} + \frac{1}{2}\lambda^2 - \frac{1}{4}e^{-4\lambda^2} + \frac{1}{4} = 3\lambda^2 + o(\lambda^3). \quad (43)$$

From Eq. (43), We can see that as the counter-rotating terms arise, the uncertainty $\Delta N$ is no longer zero, which indicates the break-down of the U(1)-symmetry.

**Conclusions**

In this work, we have eliminated the counter-rotating terms approximately and obtained the analytical energy spectrum and wavefunctions. In the weak counter-rotating coupling limit we find out that the counter-rotating terms can be considered as the shifts to the parameters of the JC model. This modification of JC model shows the validness of RWA on the assumption of near-resonance and relatively weak coupling. Finally, we derive the analytical expressions of the Bloch-Siegert shift and the uncertainties of the number of polaritons. The results show the break-down of the U(1)-symmetry and the deviation from the JC model.

522-527 (1940).

**Acknowledgments**

This work is supported by the National Natural Science Foundation of China (Grant No.11174024), and the Open Project Program of State Key Laboratory of Low-Dimensional Quantum Physics (Tsinghua University) grants Nos. KF201407, also supported by the Open Project Program of State Key Laboratory of Theoretical Physics, Institute of Theoretical Physics, Chinese Academy of Sciences, China (No.Y4KF201CJ1) and Beijing Higher Education Young Elite Teacher Project）YETP 1141.

**Author contributions**

G.Z. conceived and designed the research. G.Z. and H.Z. performed analysis and wrote the manuscript. H.Z. prepared all the figures.

**Additional information**

Competing financial interests: The authors declare no competing financial interests.